\documentclass[letterpaper,twocolumn,prb,aps,showpacs,floatfix, 
superscriptaddress]{revtex4-2}
\usepackage[latin1]{inputenc}
\usepackage{bm}
\usepackage[usenames]{color}
\usepackage{multirow}
\usepackage{amssymb}
\usepackage{amsbsy}
\usepackage{amsmath}
\usepackage{stmaryrd}
\usepackage{graphicx}
\usepackage{epsfig}
\usepackage{placeins}
\usepackage{bbold}
\usepackage{bbm}
\usepackage{braket}
\usepackage[colorlinks,linkcolor=blue,citecolor=blue,urlcolor=blue]{hyperref}

\usepackage{scalerel}
\usepackage{tikz}
\usetikzlibrary{calc}
\usetikzlibrary{patterns}
\usetikzlibrary{svg.path}
\definecolor{orcidlogocol}{HTML}{A6CE39}
\tikzset{
  orcidlogo/.pic={
    \fill[orcidlogocol] 
svg{M256,128c0,70.7-57.3,128-128,128C57.3,256,0,198.7,0,128C0,57.3,57.3,0,128,
0C198.7,0,256,57.3,256,128z};
    \fill[white] svg{M86.3,186.2H70.9V79.1h15.4v48.4V186.2z}
                 
svg{M108.9,79.1h41.6c39.6,0,57,28.3,57,53.6c0,27.5-21.5,53.6-56.8,
53.6h-41.8V79.1z 
M124.3,172.4h24.5c34.9,0,42.9-26.5,
42.9-39.7c0-21.5-13.7-39.7-43.7-39.7h-23.7V172.4z}
                 
svg{M88.7,56.8c0,5.5-4.5,10.1-10.1,10.1c-5.6,0-10.1-4.6-10.1-10.1c0-5.6,4.5-10.1
,10.1-10.1C84.2,46.7,88.7,51.3,88.7,56.8z};
  }
}

\newcommand\orcid[1]{\!%
  \href{https://orcid.org/#1}{%
    \mbox{%
      \scaleto{%
        \begin{tikzpicture}[yscale=-1,transform shape]
          \pic{orcidlogo};
        \end{tikzpicture}
      }{8pt}%
    }%
  }%
}

\makeatletter

\begin{document}
\title{Transport and integrability-breaking in non-Hermitian many-body quantum systems}

\author{Dylan E. Mahoney~\orcid{0000-0002-8327-7341}}
\affiliation{Department of Physics, Stanford University, Stanford, CA 94305, 
USA}

\author{Jonas Richter~\orcid{0000-0003-2184-5275}}
\affiliation{Department of Physics, Stanford University, Stanford, CA 94305, 
USA}
\affiliation{Institut f\"ur Theoretische Physik, Leibniz 
Universit\"at Hannover, 30167 Hannover, Germany}

\date{\today}

\begin{abstract}

Describing open quantum systems in terms of effective non-Hermitian Hamiltonians gives rise to non-unitary time evolution. In this paper, we study the impact of non-unitary dynamics on the emergent hydrodynamics in quantum systems with a global conservation law. To this end, we demonstrate how linear-response correlation functions can be generalized and interpreted in the case of non-Hermitian systems. Moreover, we show that dynamical quantum typicality provides an efficient numerical approach to evaluate such correlation functions, even though the non-unitary dynamics leads to subtleties that are absent in the Hermitian case. As a point of reference for our analysis, we consider the Hermitian spin-$1/2$ XXZ chain, whose high-temperature transport properties have been characterized extensively in recent years. Here, we explore the resulting hydrodynamics for different non-Hermitian perturbations of the XXZ chain. We also discuss the role of integrability by studying the complex energy-level statistics of the non-Hermitian quantum models.   
\end{abstract}

\maketitle


\section{Introduction}

The relaxation of many-body quantum systems towards thermal equilibrium is a topic which has attracted much interest over the last few decades \cite{dAlessio2016quantum, NandkishoreHuse_review, Gogolin_2016, Borgonovi_2016}. In the presence of a global conservation law, e.g., energy, particle number, or magnetization, the long-time dynamics of such systems are governed by an effective hydrodynamic description that arises from the underlying microscopic equations of motion \cite{Khemani_Hydro_PRX2018, Rakovszky2018, Bertini_2021}. In chaotic systems, the emerging hydrodynamics are usually expected to be diffusive \cite{Khemani_Hydro_PRX2018, Rakovszky2018, Bertini_2021, Bohrdt2017, Lux2014, Richter2019f}. In contrast, in the case of integrable systems, the extensive set of constants of motions typically leads to ballistic or superdiffusive transport \cite{Heidrich-Meisner2007, Gopalakrishnan_2023}, which can also be understood within the framework of generalized hydrodynamics \cite{Bertini2016, Castro-Alvaredo2016}. Exploring the emergence of different types of transport is an active area of research both on the experimental side, especially in cold-atom and trapped-ion platforms \cite{Wei2022, Joshi_2022, Jepsen_2020}, but also in more traditional solid-state settings \cite{Hess_2019, Scheie_2021}, as well as on the theoretical side, where sophisticated numerical techniques are being developed \cite{White2018, Ye2020, Rakovszky2022, Klein_Kvorning_2022}. 

While the time evolution of an ideally isolated quantum system is unitary and described by the Schr\"odinger equation, perfect isolation from an environment is not always realistic. The dynamics of the actual open system 
might then be described in terms of, e.g., quantum master equations or by considering suitable non-Hermitian Hamiltonians \cite{Breuer2007, Daley_2014, Ashida_2020, Bender_1998}. Not least spurred by the improved experimental control over non-Hermitian systems \cite{Cao_2015, Regensburger_2012, Zhang_2016, Feng_2017, R_ter_2010}, various fascinating aspects have been explored in recent years, including the non-Hermitian skin effect \cite{Yao_2018, Herviou_2019, Helbig_2020}, exceptional points and generalizations of topological phases \cite{Heiss_2012, Gong_2018, Leykam_2017, Kunst_2018, Luitz_2019, Bergholtz_2021}, quantum chaos \cite{Akemann_2019, Li_2021, Denisov_2019, S__2020, Garc_a_Garc_a_2022, Wang_2020, Can_2019, Cornelius_2022, Kawabata_2023, Sharma2024}, eigenstate thermalization \cite{Roy2023, Cipolloni2023}, and many-body localization \cite{Hamazaki_2019, Mak2023, Zhai_2020, Suthar_2022}. The study of non-Hermitian systems is interesting also in a broader context, as the non-unitary dynamics provides a framework to study new phenomena and realize novel out-of-equilibrium phases of matter \cite{skinnerMeasurementInducedPhaseTransitions2019, liQuantumZenoEffect2018, potter2021entanglement, Sieberer2023,Tibor_2023}.  

In this paper, we explore how non-unitary time evolution, governed by non-Hermitian Hamiltonians, affects the emerging hydrodynamics in quantum systems with a conserved quantity. While there is a long history of studying transport in boundary-driven systems, where the system-bath setup is modelled by a Lindblad master equation \cite{Bertini_2021, Znidaric2011}, studying transport directly from the point of view of non-Hermitian Hamiltonians has received less attention. As a convenient starting point, we consider the integrable and Hermitian spin-$1/2$ XXZ chain, the high-temperature transport properties of which are well established in the literature \cite{Bulchandani_2021, Gopalakrishnan_2023, Bertini_2021}. In particular, we will study different non-Hermitian deformations of the XXZ model, including an interacting spin-chain version of the Hatano-Nelson model with asymmetric hopping amplitudes, as well as a disordered spin chain with non-Hermitian random-field terms.   

Transport in many-body quantum systems is commonly studied in terms of time-dependent linear-response correlation functions \cite{Bertini_2021}. We here proceed in an analogous way for the non-Hermitian setting and propose a generalization of such correlation functions to systems with non-unitary dynamics \cite{Brody_2012}. This generalization has an appealing experimental interpretation and connects to standard definitions of non-unitary time evolution in quantum systems. In order to numerically evaluate these correlation functions, we moreover demonstrate that the concept of dynamical quantum typicality (DQT) is applicable also in the case of non-Hermitian quantum systems. In particular, we show that on the time scales where the dynamics exhibit hydrodynamic behavior, DQT yields accurate results if the system sizes are sufficiently large. 

The non-Hermitian perturbations considered in this paper are found to affect the transport properties of the original XXZ chain in different ways, with relaxation becoming faster or slower depending on the model. We also study the role of integrability and find that while certain non-Hermitian perturbations leave the integrability of the XXZ chain intact, others induce the emergence of chaos and random-matrix energy level statistics. Interestingly, our results suggest that there exist short-range non-Hermitian quantum systems which support faster than diffusive, or even ballistic, transport, despite being chaotic and nonintegrable. 

The rest of this paper is structured as follows. We define the models and observables in Sec.\ \ref{Sec::Model}, including 
 the complex eigenvalue-gap ratio as an indicator of chaos in non-Hermitian systems \cite{S__2020}. We also discuss our generalization of dynamical correlation functions in the case of non-Hermitian systems. In Sec.\ \ref{Sec::Method}, we then highlight DQT as a means to simulate such correlation functions for systems sizes beyond the range of full exact diagonalization (ED). Our numerical results are presented in Sec.\ \ref{Sec::Results}, where we explore integrability-breaking and transport properties for different non-Hermitian Hamiltonians. We also discuss the challenges that occur when applying DQT for systems undergoing non-unitary time evolution. We summarize and conclude in Sec.\ \ref{Sec::Conclu}.


\section{Models and Observables}\label{Sec::Model}

We study an interacting version of the Hatano-Nelson model written in terms of spin-$1/2$ operators \cite{Hatano_1996, Hamazaki_2019, Zhang_2022, Panda_2020, Mak2023, Heu_en_2021},  
\begin{align}\label{Eq::HN}
    H = \sum_{\ell=1}^L &\frac{1}{2}\left(e^g S_\ell^+ S_{\ell+1}^- + e^{-g} S_{\ell}^- S_{\ell+1}^+ \right) \\ &+ \Delta S_\ell^z S_{\ell+1}^z + \Delta_2 S_\ell^z S_{\ell+2}^z\ \nonumber , 
\end{align}
where $S_\ell^\pm = S_\ell^x \pm i S_\ell^y$, $\Delta \geq 0$ ($\Delta_2 \geq 0$) controls the strength of (next-)nearest neighbor interactions, $L$ is the system size, and we consider periodic boundary conditions.  While Eq.\ \eqref{Eq::HN} is Hermitian at $g = 0$, a finite $g \neq 0$ leads to an asymmetry in the hopping amplitudes such that the model becomes non-Hermitian, $H^\dagger \neq H$. 

 At $g = 0$ and $\Delta_2 = 0$, $H$ reduces to the paradigmatic integrable spin-$1/2$ XXZ chain. While spin transport is ballistic in the XXZ chain for $\Delta < 1$ with a finite Drude weight, numerical evidence indicates that it is diffusive for $\Delta > 1$ \cite{Bertini_2021}. Moreover, at the isotropic point $\Delta = 1$, spin transport appears to be superdiffusive with certain features being described by Karder-Parisi-Zhang universality \cite{Ljubotina_2019}. For finite $\Delta_2$, the XXZ chain becomes nonintegrable and transport appears diffusive for all choices of $\Delta \neq 0$ and $\Delta_2 \neq 0$ \cite{Richter_2018}. In this paper, we study how integrability and transport characteristics change when turning to the non-Hermitian system with $g \neq 0$. 

The nonreciprocal hopping terms in Eq.\ \eqref{Eq::HN} are reminiscent of the dynamical rules in asymmetric simple exclusion processes known from nonequilibrium statistical mechanics \cite{Derrida_1998,Nakerst_2024}. Indeed, mappings between such classical hopping models and non-Hermitian Hamiltonians have been considered \cite{de_Gier_2005}. Moreover, analogous to the standard XXZ chain, the asymmetric variant in Eq.\ \eqref{Eq::HN} with $\Delta_2 = 0$ is known to be Bethe-Ansatz integrable \cite{Golinelli_2006} (see also a similarly integrable non-Hermitian Bose-Hubbard model in \cite{Zheng_2024}). We confirm its integrablity below in terms of the level-spacing statistics.     

More recently, the Hamiltonian \eqref{Eq::HN} has been studied in different contexts, including the addition of quenched disorder which can induce non-Hermitian many-body localization \cite{Hamazaki_2019, Mak2023}. In the context of transport, let us note that the asymmetric hopping terms can be obtained from $\cosh(g) H_{XY} + i \sinh(g) J$, where $H_{XY} = (1/2)\sum_\ell S_\ell^+ S_{\ell+1}^- + S_\ell^- S_{\ell+1}^+$, and $J = (i/2)\sum_\ell S_\ell^+ S_{\ell+1}^- - S_\ell^- S_{\ell+1}^+$ is the spin-current operator. The non-Hermiticity in Eq.\ \eqref{Eq::HN} can thus be interpreted as an external driving by the current $J$ \cite{Panda_2020}. 

While our focus will be on the model with asymmetric hopping in Eq.\ \eqref{Eq::HN}, we will also study an XXZ chain perturbed by imaginary random on-site potentials \cite{Roccati2023, Chen_2023}, 
\begin{align}\label{Eq::HDisorder}
    H = \sum_{\ell = 1}^L S_\ell^x S_{\ell+1}^x + S_{\ell}^y S_{\ell+1}^y + S_\ell^z S_{\ell+1}^z - i h_\ell n_\ell\ , 
\end{align}
where $n_\ell = S_\ell^z + \frac{1}{2}$, and the $h_\ell \in [0,W]$ are drawn at random from a uniform distribution with $W$ setting the disorder strength. Equation \eqref{Eq::HDisorder} closely resembles the disordered XXZ chain that has become the prototypical model to study the MBL transition \cite{Abanin_MBL}. In our case, however, the disorder in Eq.\ \eqref{Eq::HDisorder} is non-Hermitian. The dynamics generated by the Hamiltonian in Eq.\ \eqref{Eq::HDisorder} can be interpreted as the no-jump trajectory in the stochastic unravelling of a Markovian open quantum system. In the trajectory approach, pure states evolve according to an effective non-Hermitian Hamiltonian \cite{Dalibard_1992}, $H_\text{eff} = H - (i/2)\sum_j \gamma_j L_j^\dagger L_j$, where $L_j$ are the jump operators occurring in the Lindblad equation, $\gamma_j > 0$, and Eq.\ \eqref{Eq::HDisorder} would correspond to $L_j \sim S_j^-$.  While at large $W$, this type of disorder might cause a localization transition \cite{Roccati2023}, we expect the non-Hermitian terms in Eq.\ \eqref{Eq::HDisorder} to favor thermalization if disorder is moderate. This expectation is substantiated by our numerical results below for $W = 1$.

\subsection{Level-spacing statistics with complex eigenvalues}

In the case of Hermitian systems with real-valued spectrum, a useful quantity to distinguish integrable from chaotic systems is the ratio of adjacent level spacings, $\text{min}(\delta_m,\delta_{m+1})/\text{max}(\delta_m,\delta_{m+1})$, where $\delta_m = E_{m+1}-E_m$ \cite{Oganesyan_2007}. 
A generalization of such level-spacing ratios to non-Hermitian systems with complex spectra has been introduced in \cite{S__2020}. In particular, given a Hamiltonian with complex eigenvalues $E_m$, one can define,
\begin{equation}\label{Eq::zm}
    z_m = \frac{E_m^\text{NN}-E_m}{E_m^\text{NNN}-E_m} \equiv \varrho_m e^{i\theta_m}\ , 
\end{equation}
where $E_m^\text{NN}$ and $E_m^\text{NNN}$ are the nearest and next-nearest neighbors of $E_m$ (i.e., the eigenvalues that minimize the euclidian distance on the complex plane). The $z_m$ are complex numbers within the unit circle ($\varrho_m \leq 1$; $-\pi \leq \theta_m \leq \pi$), for which we can study the full distribution $P_{z}(\varrho,\theta)$, as well as average values, e.g., $\langle \varrho \rangle$ and $\langle \cos \theta \rangle$. In particular, for integrable systems, the $z_m$ are expected to be uniformly distributed such that $\langle \varrho \rangle = 2/3$ and $\langle \cos \theta \rangle = 0$. In contrast, for nonintegrable non-Hermitian systems, the distribution of $z_m$ is anisotropic, which will lead to different $\langle \varrho \rangle$ and $\langle \cos \theta \rangle$ \cite{S__2020}. 

\subsection{Dynamical correlation functions}

A common approach to study transport properties in Hermitian many-body quantum systems is to consider dynamical correlation functions (often at formally infinite temperature), $C(r,t)  = \text{tr}[S_{\ell+r}^z(t) S_\ell^z \rho_\infty]$,
where $S_\ell^z(t)$ is the time-evolved operator in the Heisenberg picture, $r$ is a distance between two sites,  and $\rho_\infty = \mathbb{1}/2^L$ is the infinite-temperature density matrix. 
In the case of non-Hermitian systems with non-unitary time evolution, one has to be careful how to meaningfully define such correlation functions, see Appendix \ref{Sec::App_Correl}. Building on previous works \cite{Brody_2012}, we define the time evolution of a non-Hermitian system prepared in some density matrix $\rho(t)$ as,
\begin{equation}\label{Eq::time_evolution_for_rho}
    \rho(t) = \frac{e^{-iHt} \rho(0) e^{iH^\dagger t}}{\text{tr}[e^{-iHt} \rho(0) e^{iH^\dagger t}]}\ , 
\end{equation}
where the time-dependent renormalization becomes trivial if $H$ is Hermitian.
Next, we use that $S_\ell^z$ commutes with the total magnetization $M^z = \sum_\ell S_\ell^z$, and we decompose both operators in terms of projectors onto their eigenspaces, i.e., 
$S_\ell^z = \sum_s s \Pi_{s}$ and $M^z = \sum_m m \Pi_m$, where $s = \pm 1/2$ and $m = -L/2,\dots,L/2$. As we discuss more generally in Appendix \ref{Sec::App_Correl}, we can then write,
\begin{equation}\label{Eq::Correl_NonHermitian}
    C(r,t) = \sum_{s,m} s\ \text{tr}[\Pi_{s,m} \rho_\infty]\langle S_{\ell+r}^z(t)\rangle_{\Pi_{s,m}}\ , 
\end{equation}
where we introduced the time-dependent expectation value,
\begin{equation}\label{Eq::Exp}
    \langle S_{\ell+r}^z(t) \rangle_{\Pi_{s,m}} = \text{tr}[S_{\ell+r}^z \rho_{s,m}(t)]\ , 
\end{equation}
which is conditional on the measurement outcome of $S_\ell^z$ and $M^z$ at $t = 0$. Namely, $\rho_{s,m}(0) = \Pi_{s,m} \rho_\infty \Pi_{s,m}/\text{tr}[\Pi_{s,m} \rho_\infty]$ is the infinite-temperature density matrix projected into a common eigenspace of $S_\ell^z$ and $M^z$ with eigenvalues $s$ and $m$. Since $\Pi_{s,m}$ is a projection, $\Pi_{s,m}^2 = \Pi_{s,m}$. Moreover, the time evolution of $\rho_{s,m}(t)$ in Eq.\ \eqref{Eq::Exp} is interpreted as in Eq.\ \eqref{Eq::time_evolution_for_rho}.

From a, say, experimental point of view, writing the correlation function $C(r,t)$ as in Eq.\ \eqref{Eq::Correl_NonHermitian} has a straightforward interpretation. Measure $S_\ell^z$ and $M^z$ at $t = 0$, measure $S_{\ell+r}^z$ at time $t$, and over many runs compute the time-dependent expectation value conditional on the measurent outcomes and corresponding probabilities at $t = 0$. 

Conceptually, the advantage of Eq.\ \eqref{Eq::Correl_NonHermitian} is that the total density, $\sum_r C(r,t)$, remains conserved over the course of the non-unitary time evolution such that the analysis of transport properties is meaningful. In standard Hermitian systems, the conservation of magnetization is already guaranteed by $[H,M^z] = 0$. In contrast, for non-Hermitian systems, the renormalization \eqref{Eq::time_evolution_for_rho} of the state during the non-unitary time evolution could cause $M^z$ to be non-conserved if one were to proceed without projecting $\rho(0)$ into a fixed $M^z$ sector; see Appendix \ref{Sec::App_Correl} for more details.

Given the correlation function $C(r,t)$, the type of transport can then be inferred from its spatiotemporal dynamics. For instance, at sufficiently long times, the on-site correlator $C(0,t)$ is expected to develop a hydrodynamic power-law tail, $C(0,t) \propto t^{-\beta}$, where (in one dimension) $\beta = 1/2$ indicates diffusion, $\beta = 1$ indicates ballistic transport, and $1/2 < \beta < 1$ ($\beta < 1/2$) signals superdiffusion (subdiffusion) \cite{Bertini_2021}.

\section{Methods}\label{Sec::Method}

\subsection{Dynamical quantum typicality}

The notion of quantum typicality describes the fact that random pure quantum states, drawn from a high-dimensional Hilbert space, can accurately represent properties of the full statistical ensemble \cite{Gemmer2004, Goldstein2006, Popescu2006, Reimann2007}. From a numerical point of view, this can be exploited to approximate equilibrium expectation values by estimating the trace over the Hilbert space using random pure states \cite{Hams2000, Elsayed2013,Monnai2014, Steinigeweg2014, Iitaka2003, Sugiura2013, Bartsch2009}, see \cite{Jin_2021, Heitmann_2020} for reviews. More concretely, taking the spatiotemporal correlation function $C(r,t)$ in Eq.\ \eqref{Eq::Correl_NonHermitian} as an example, we can approximate
it by the pure-state estimate, 
\begin{equation}\label{Eq::TypCrt}
   C_{\psi}(r,t) =  \sum_{s,m} s\ \frac{\text{Tr}[\Pi_{s,m}]}{2^L}\bra{\psi_{s,m}(t)}S_{\ell+r}^z \ket{\psi_{s,m}(t)}\ , 
\end{equation}
where we have introduced the projected state $\ket{\psi_{s,m}}$, 
\begin{equation}\label{Eq::RandomState}
    \ket{\psi_{s,m}} = \frac{\Pi_{s,m} \ket{\psi}}{\sqrt{\braket{\psi|\Pi_{s,m}|\psi}}}\ ,
\end{equation}
and $\ket{\psi} = \sum_{k=1}^{2^L} c_k \ket{k}$, is a random state in the computational basis with the complex coefficients $c_k$ drawn from a Gaussian distribution with zero mean, and then normalized so that $|| \ket{\psi}|| = 1$. (See Appendix \ref{Sec::App:Typ} for more details.) Thus, $\ket{\psi_{s,m}}$ is a random state in a subspace with fixed eigenvalue $s$ of $S_\ell^z$ and $m$ of $M^z$.

The numerical advantage of DQT stems from the fact that one does not need to treat the full density matrix $\rho(t)$, but only has to deal with pure states instead. Crucially, for local Hamiltonians, the time evolution of pure states can be achieved by efficient sparse matrix techniques without full exact diagonalization. As a consequence, system sizes beyond the range of ED can be studied and we here present results for $C(r,t)$ up to $L \leq 24$.  
Analogous to Eq.\ \eqref{Eq::time_evolution_for_rho}, we explicitly preserve the state's norm over the course of the time evolution governed by the non-Hermitian $H$, i.e., 
\begin{equation}\label{Eq::Psit}
    \ket{\psi_{s,m}(t)} = \frac{e^{-iHt} \ket{\psi_{s,m}(0)}}{|| e^{-iHt} \ket{\psi_{s,m}(0)}||  }\ ,  
\end{equation}
where, in practice, we will actually evolve the state by a (sufficiently small) discrete time step using a Runge-Kutta scheme, $\ket{\psi_{s,m}(t+\delta t)} = e^{-iH\delta t} \ket{\psi_{s,m}(t)}$, and normalize after each such time step. 
  
DQT has been applied extensively as a useful numerical tool to study Hermitian many-body quantum systems. It relies on the largeness of the Hilbert space such that the accuracy of the random-state approximation $C_\psi(r,t) \approx C(r,t)$ improves exponentially with increasing system size \cite{Jin_2021}. In practice, a single realization of the random state $\ket{\psi}$ is often enough to obtain results with negligible statistical errors. 

In case of non-Hermitian $H$, however, the time evolution in Eq.\ \eqref{Eq::Psit} will enhance (suppress) the relative weight of the wave function's components. Namely, in the long-time limit, $\ket{\psi_{s,m}(t)}$ will converge towards the eigenstate of $H$ that belongs to the eigenvalue with the largest imaginary part (or a superposition of such eigenstates if that eigenvalue has a degeneracy). Thus, with growing $t$, fewer eigenstates will contribute to $\ket{\psi_{s,m}(t)}$ such that the typicality approximation is expected to become less accurate.  

We can study the accuracy of DQT by considering the relative variance \cite{Schiulaz_2020}, 
\begin{equation}\label{Eq::Rt}
    R(t) = \frac{\overline{C_{\psi}(r,t)^2} - \overline{C_{\psi}(r,t)}^2}{ \overline{C_{\psi}(r,t)}^2}\ , 
\end{equation}
where the overline indicates averaging over random realizations of the state $\ket{\psi}$. Self-averaging behavior is indicated by an $R(t)$ that decreases with increasing system size \cite{Schiulaz_2020}. In the Hermitian case, $R(t)$ decreases exponentially with $L$, highlighting that only a single instance of $\ket{\psi}$ is required if $L$ is sufficiently large \cite{Richter_2020}.


\section{Results}\label{Sec::Results}

\subsection{Breaking of integrability}\label{Sec::Res::Inte}

We study the impact of $g > 0$ and $\Delta_2 > 0$ on the integrability of the Hamiltonian in Eq.\ \eqref{Eq::HN} by considering its energy-level statistics.
In order to avoid mixing of eigenvalues with different quantum numbers, our results are obtained in the symmetry subspace with lattice momentum $q = 2\pi/L$ and total magnetization $M^z = 1$. This choice also resolves potential spin-flip or reflection symmetries of $H$.

\begin{figure}[tb]
\centering
\includegraphics[width=\columnwidth]{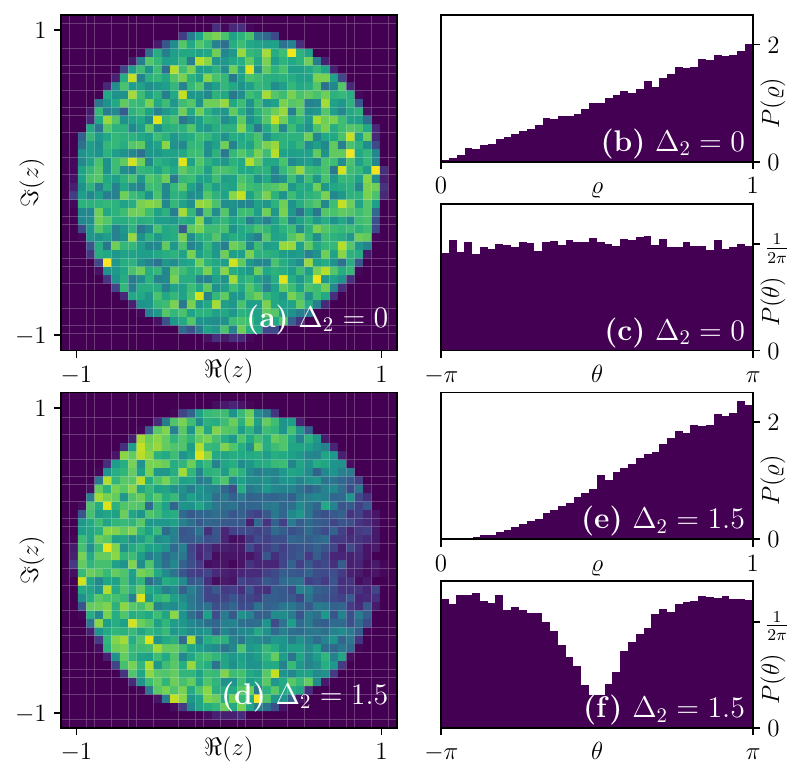}
\caption{Level-spacing statistics for the Hatano-Nelson model with $L = 22$, $\Delta = 1.5$, and $g = 0.2$.  {\bf [(a) - (c)]} Distribution of  $z = \varrho e^{i \theta}$ on the complex plane  [Eq.~\eqref{Eq::zm}], as well as marginal distributions $P(\varrho)$ and $P(\theta)$, obtained for $\Delta_2 = 0$. {\bf [(d) - (f)]} Analogous data, but now for $\Delta_2 = 1.5$.}
\label{Fig::LevelStatHN_1}
\end{figure}

In Fig.\ \ref{Fig::LevelStatHN_1}, we study the distribution of level-spacing ratios $z$, cf.\ Eq.\ \eqref{Eq::zm}, for $L = 22$, $\Delta = 1.5$, and fixed non-Hermitian hopping asymmetry $g = 0.2$. In Fig.\ \ref{Fig::LevelStatHN_1}~(a) - (c), data are shown for the nearest-neighbor chain with $\Delta_2 = 0$. We find that the $z$ are approximately homogeneously distributed over the unit circle with the corresponding marginal distributions $P(\varrho) \approx 2\varrho$ and $P(\theta) \approx 1/(2\pi)$. Such behavior is expected for integrable non-Hermitian systems \cite{S__2020}. Thus, while a finite $g$ breaks the Hamiltonian's Hermiticity, $g > 0$ is not sufficient to lift the original integrability of the XXZ chain \cite{Golinelli_2006, de_Gier_2005, S__2020}. 

The picture is clearly different when considering next-nearest neighbor interactions with $\Delta_2 = 1.5$, see Fig.\ \ref{Fig::LevelStatHN_1}~(d) - (f). In particular, we find that spacing ratios $z$ with small $\varrho \to 0$ are less likely to occur. Moreover, the distribution becomes $\theta$-dependent with suppressed weight around $\theta = 0$. This indicates that, analogous to the Hermitian XXZ chain \cite{Richter_2018}, a finite $\Delta_2 > 0$ will make the model nonintegrable. 

\begin{figure}[tb]
\centering
\includegraphics[width=\columnwidth]{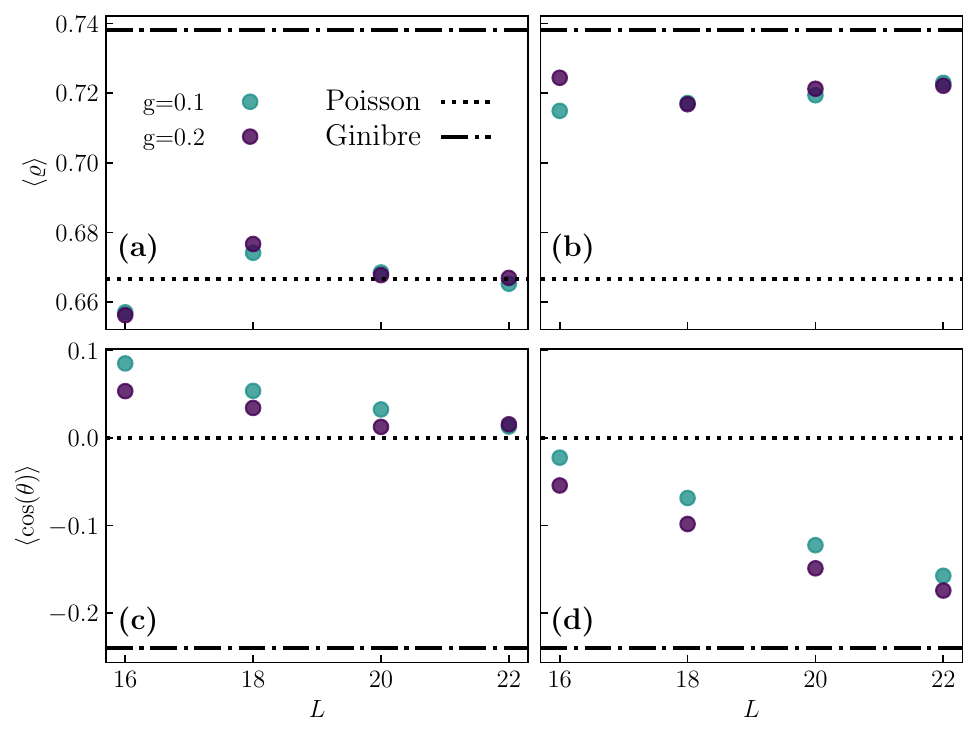}
\caption{Finite-size scaling of average values $\langle \varrho\rangle$ and $\langle \cos \theta \rangle$ for $g = 0.1,0.2$ and $L = 16,18,20,22$. Data are compared to the predictions of the Poisson (dotted) and Ginibre (dashed-dotted) ensembles \cite{S__2020}. {\bf [(a), (b)]} $\langle \varrho\rangle$ for $\Delta_2 = 0$ and $\Delta_2 = 1.5$. {\bf [(c), (d)]} $\langle \cos \theta\rangle$ for $\Delta_2 = 0$ and $\Delta_2 = 1.5$. We have $\Delta = 1.5$ in all cases.}
\label{Fig::LevelStatHN_2}
\end{figure}

From the distributions $P(\varrho)$ and $P(\theta)$ in Fig.\ \ref{Fig::LevelStatHN_1}, we obtain the averages $\langle \varrho \rangle$ and $\langle \cos \theta \rangle$, which are shown in Fig.\ \ref{Fig::LevelStatHN_2} versus system size $L$. While for $\Delta_2 = 0$, $\langle \varrho \rangle$ and $\langle \cos \theta \rangle$ are close to the expected values of the integrable Poisson distribution, we find that the data for $\Delta_2 = 1.5$ approaches with increasing $L$ the prediction of the nonintegrable Ginibre ensemble as described in \cite{S__2020}. 
These results substantiate our findings from Fig.\ \ref{Fig::LevelStatHN_1} that the non-Hermitian perturbation $g > 0$ is not sufficient to break integrability, whereas the next-nearest neighbor interaction $\Delta_2 > 0$ causes the model to become chaotic. 

In Appendix \ref{Sec::App_LSDisorder}, we present additional results for the level-spacing statistics of the XXZ chain perturbed by random imaginary fields, cf.\ Eq.\ \eqref{Eq::HDisorder}. In contrast to the non-Hermitian hopping asymmetry $g \neq 0$ [Fig.\ \ref{Fig::LevelStatHN_1}~(a)], the results in Appendix \ref{Sec::App_LSDisorder} suggest that the XXZ chain indeed becomes nonintegrable in the presence of non-Hermitian disorder \cite{Roccati2023}.

\subsection{DQT in non-Hermitian systems}\label{Sec::Res::Typ}

\begin{figure}[tb]
\centering
\includegraphics[width=\columnwidth]{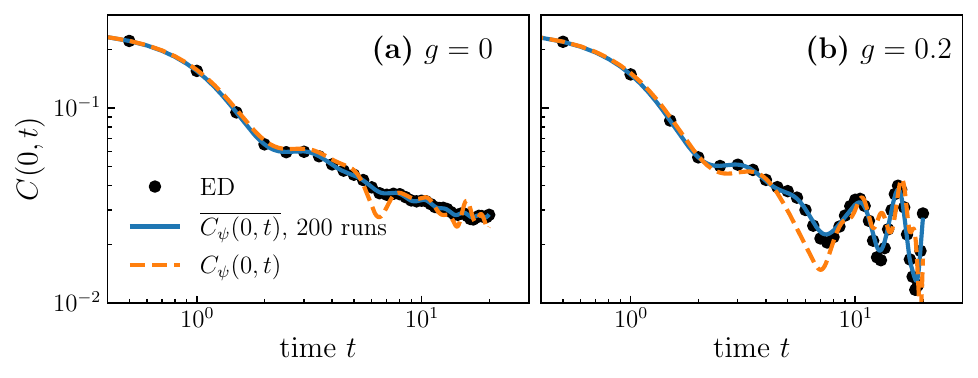}
\caption{Comparison between dynamical quantum typicality and exact diagonalization for system size $L = 10$. {\bf [(a), (b)]} Spin autocorrelation function $C(0,t)$ for $g=0$ and $g = 0.2$. DQT data (solid curves), averaged over $200$ random $\ket{\psi}$, agree convincingly with ED (symbols). 
DQT data obtained from a single random state (dashed curves) show clear deviations from the average especially in the non-Hermitian case. We have $\Delta = 1.5$ and $\Delta_2 = 0$ in all cases.
}
\label{Fig::Typicality_1}
\end{figure}

Let us now analyze the accuracy of dynamical quantum typicality applied to systems with non-unitary time evolution. To this end, in Fig.\ \ref{Fig::Typicality_1}, we present a comparison between DQT and ED using a small system size $L = 10$. We consider the on-site spin correlation function $C(0,t)$ and study both Hermitian ($g = 0$) and non-Hermitian ($g = 0.2$) systems. For a single exemplary realization of the random state $\ket{\psi}$, we find that the pure-state estimate $C_\psi(0,t)$ exhibits clear deviations from the exact result.

These fluctuations are expected for $L = 10$ where the Hilbert-space dimension is still not big enough to suppress the statistical error sufficiently. Moreover, while the fluctuations of $C_\psi(0,t)$ in the Hermitian case remain fairly controlled for all times shown here [Fig.\ \ref{Fig::Typicality_1}~(a)], they become more pronounced with increasing $t$ for $g = 0.2$ [Fig.\ \ref{Fig::Typicality_1}~(b)]. This phenomenon is discussed in more detail in the context of Fig.\ \ref{Fig::Typicality_2} below. 

As mentioned in Sec.\ \ref{Sec::Method}, the accuracy of the pure-state approximation \eqref{Eq::TypCrt} can be improved by averaging over multiple random realization of $\ket{\psi}$. Indeed, as shown in Fig.\ \ref{Fig::Typicality_1}, the averaged correlation function $\overline{C_\psi(0,t)}$ (here obtained from $200$ independent runs) agrees perfectly with the ED data. Furthermore, in Appendix \ref{Sec::App::CurrCurr}, we show that such a convincing agreement between the random-state approach and ED can be obtained for other classes of correlation functions as well, i.e., current-current correlation functions also relevant in the context of transport. 

Strictly speaking, quantum typicality refers to the fact that (for large Hilbert spaces) the pure-state estimate $C_\psi(r,t)$ is close to the exact result such that no averaging is required. 
To study this issue in more detail, we show 
in Fig.\ \ref{Fig::Typicality_2} the averaged correlation function $\overline{C_\psi(0,t)}$ for $g = 0$ and $g = 0.2$ [Fig.\ \ref{Fig::Typicality_2}~(a), (b)] together with their corresponding relative variance $R(t)$ [Fig.\ \ref{Fig::Typicality_2}~(c), (d)]. Plotting data for different system sizes $L$, we find that $R(t)$ decreases exponentially with increasing $L$ if $H$ is Hermitian \cite{Richter_2020}. In other words, choosing a single random $\ket{\psi}$ will yield results very close to the ensemble average $\overline{C(0,t)}$ in the Hermitian case. 

In the non-Hermitian case, we find that $R(t)$ also decreases with $L$ on short to intermediate time scales. However, $R(t)$ becomes essentially independent of $L$ at longer times, where the correlation function $C(0,t)$ has approximately reached its long-time value $C(0,t\to \infty) \to 1/(4L)$. Thus, at these time scales, self-averaging and typicality are absent such that averaging over multiple $\ket{\psi}$ is still required at larger $L$. 

Eventually, it is instructive to connect the behavior of $R(t)$ to the inverse-participation ratio of the state $\ket{\psi(t)}$,
\begin{equation}\label{Eq::IPR}
    I(t) = \frac{1}{\left(\sum_{m} |\braket{\phi_m|\psi(t)}|^2\right)^2}\sum_{n} |\braket{\phi_n|\psi(t)}|^4\ ,
\end{equation}
which measures how extended $\ket{\psi(t)}$ is in a certain basis. Here, we will choose the $\ket{\phi_n}$ as the right eigenvectors of $H$ in a fixed symmetry subspace of dimension ${\cal N}$. While it is possible to choose bases of left and right eigenvectors for a non-Hermitian Hamiltonian which are biorthogonal, it is not possible to simultaneously make every right eigenvector normalized and every left eigenvector normalized \cite{Ashida_2020}. In Eq.\ \eqref{Eq::IPR} we take each right eigenvector to be normalized. For Hermitian systems, Eq.\ \eqref{Eq::IPR} reduces to the well-known expression, 
\begin{equation}\label{Eq::Hermitian_IPR}
    I(t) = \sum_{n} |\braket{\phi_n|\psi(t)}|^4\ .
\end{equation}
An interpretation of Eq.\ \eqref{Eq::Hermitian_IPR} in terms of the outcomes of a particular measurement process is given in Appendix \ref{Sec::App_IPR} such that Eq.\ \eqref{Eq::IPR} becomes a natural extension of Eq.\ \eqref{Eq::Hermitian_IPR} to non-Hermitian systems.

For a Haar-random initial state $\ket{\psi}$, we expect it to be fully extended in any given basis such that $|\braket{\phi_n|\psi(t)}|^4 \sim 1/{\cal N}^2$ and $I(0) \sim 1/{\cal N}$ \cite{Schiulaz_2020}. 

In the Hermitian case [Fig.\ \ref{Fig::Typicality_2}~(e)], the dynamics of $I(t)$ are trivial. The unitary time evolution merely leads to different phases of the eigenstates such that $I(t)$ remains constant. In contrast, for non-Hermitian systems, the spectrum of $H$ is complex in general. The imaginary parts of the eigenvalues, combined with the renormalization of $\ket{\psi(t)}$ in Eq.\ \eqref{Eq::Psit}, will suppress or enhance the contributions of certain eigenstates \cite{Panda_2020}. Indeed, as shown in Fig.\ \ref{Fig::Typicality_2}~(f), we find that $I(t)$ starts growing significantly for $t\gtrsim 5$. Thus, the random state $\ket{\psi_{s,m}(t)}$ in Eq.\ \eqref{Eq::RandomState} becomes less extended (i.e., less typical) with increasing $t$. As a consequence, the statistical error of the pure-state estimate $C_\psi(0,t)$, which relies on the randomness of $\ket{\psi_{s,m}(t)}$, is expected to increase. This explains that the relative variance $R(t)$ in Fig.\ \ref{Fig::Typicality_2}~(d) ceases to decay with $L$ at long times.   

The analysis in Fig.\ \ref{Fig::Typicality_2} suggests that DQT is less useful in systems with non-unitary time evolution. Nevertheless, we have demonstrated in Fig.\ \ref{Fig::Typicality_1} that highly accurate results, which agree with ED, can certainly be obtained if the pure-state data are averaged over sufficiently many random $\ket{\psi}$. Moreover, in practice, we are not interested in the long-time regime where the correlation function has completely decayed and self-averaging breaks down. Rather, our focus will be on the intermediate time scales for which $C(r,t)$ exhibits hydrodynamic behavior. On these time scales, $R(t)$ still decreases with $L$ such that the accuracy of DQT is expected to improve with increasing system size. We will confirm this expectation in the next section, where we discuss transport properties based on DQT data with system sizes up to $L = 24$.
\begin{figure}[tb]
\centering
\includegraphics[width=\columnwidth]{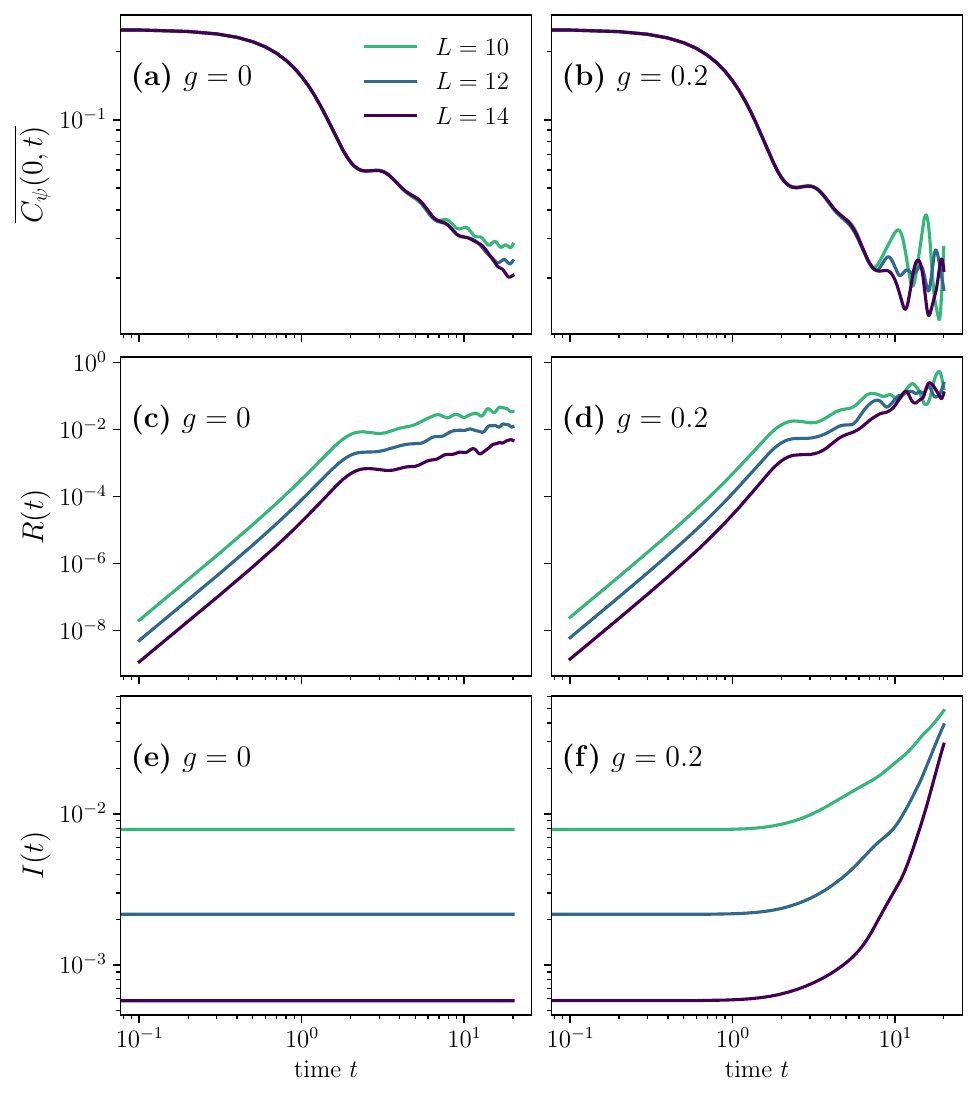}
\caption{{\bf [(a), (b)]} $\overline{C_\psi(0,t)}$ obtained by averaging over $256$ random states for system sizes $L = 10,12,14$ with $g = 0$ and $g = 0.2$. In all panels we have $\Delta = 1.5$ and $\Delta_2 = 0$. {\bf [(c), (d)]} Corresponding relative variances $R(t)$, cf.\ Eq.\ \eqref{Eq::Rt}. {\bf [(e), (f)]} Averaged inverse participation ratio $I(t)$, as defined in \eqref{Eq::IPR}, of the state $\ket{\psi_{s,m}(t)}$ for the subspace with $m = 0$.}
\label{Fig::Typicality_2}
\end{figure}

\subsection{Transport}\label{Sec::Res::Trans}

\subsubsection{Asymmetric XXZ chain}

\begin{figure}[tb]
\centering
\includegraphics[width=\columnwidth]{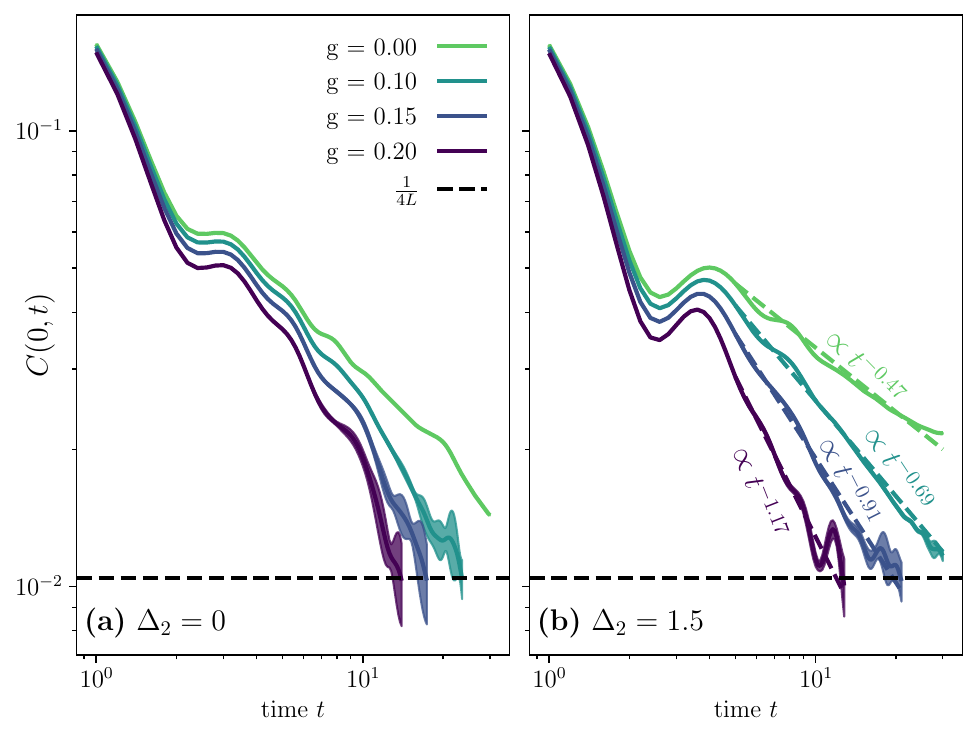}
\caption{Spin-Spin autocorrelation function $C(0,t)$ for $L = 24$ and $g = 0,0.1,0.15,0.2$, obtained using DQT. Each curve is obtained by averaging over $6$ random-state realizations, and the area within $\pm2$ standard errors of the mean (assuming Gaussian errors) is shaded. {\bf(a)} $\Delta = 1.5$ and $\Delta_2 = 0$. {\bf(b)} $\Delta = 1.5$ and $\Delta_2 = 1.5$. The dashed lines are power-law fits (beginning at $t = 5$) to the hydrodynamic tail. The horizontal line indicates the equilibrium long-time value of $C(0,t)$. For visual clarity, curves are not shown after the average of $C(0,t)$ reaches its equilibrium long-time value.}
\label{Fig::Transport_1}
\end{figure}

We now turn to the transport properties of the Hatano-Nelson model \eqref{Eq::HN}. The on-site spin-spin correlation function $C(0,t)$ is shown in Fig.\ \ref{Fig::Transport_1} for an integrable ($\Delta = 1.5$, $\Delta_2 = 0$) and a nonintegrable ($\Delta = \Delta_2 = 1.5$) parameter choice with fixed system size $L = 24$. In both cases, we show data for the Hermitian $g = 0$ chain as well as for finite non-Hermitian hopping asymmetry $g = 0.1,0.15,0.2$. 

For concreteness, we especially focus the nonintegrable model in the following [Fig.\ \ref{Fig::Transport_1}~(b)], for which the numerical data is somewhat cleaner and easier to interpret. The overall phenomenology, however, also applies to the $\Delta_2 = 0$ model. As a point of reference, we observe a hydrodynamic power-law tail consistent with $C(0,t) \propto t^{-1/2}$ for the Hermitian $g = 0$ case, indicating diffusive transport as expected in the easy-axis regime of the (nonintegrable) XXZ model \cite{Bertini_2021}. Moreover, we find that adding a finite hopping-asymmetry has a striking effect on the resulting dynamics, with $C(0,t)$ decaying significantly faster. Specifically, on the finite time and length scales numerically available to us, the dynamics at $g = 0.2$ are consistent with $C(0,t) \propto t^{-\beta}$ with $\beta \gtrsim 1$, which suggests that transport becomes ballistic in the non-Hermitian model. In particular, these signatures of ballistic transport represent an exception to the common belief that chaotic systems with short-range interactions exhibit conventional diffusion.

We note that the data in Fig.\ \ref{Fig::Transport_1} for $L = 24$ are obtained using the pure-state approach \eqref{Eq::TypCrt} by averaging over just a few ($\leq 10$) random pure states. We find that this yields sufficiently small statistical fluctuations in the intermediate time regime where $C(0,t)$ decays as a power law. This demonstrates that DQT indeed provides a useful numerical tool to study the dynamics of non-Hermitian many-body quantum systems. 

\begin{figure}[tb]
\centering
\includegraphics[width=\columnwidth]{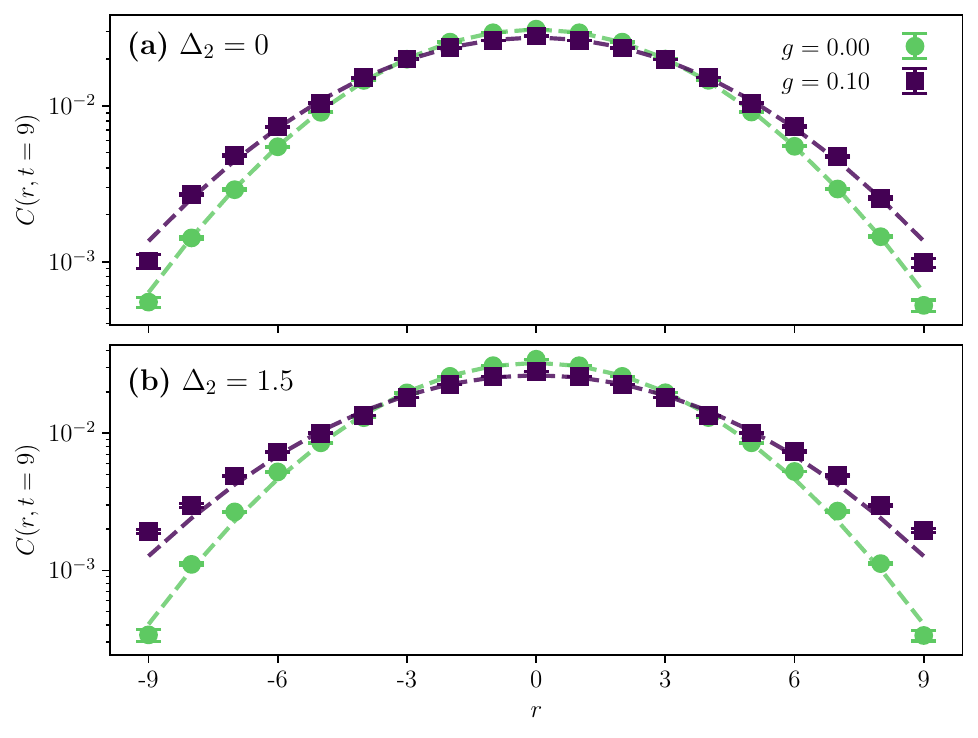}
\caption{$C(r,t)$ for $g = 0,0.1$ at fixed time $t = 9$. {\bf(a)} $\Delta = 1.5$ and $\Delta_2 = 0$. {\bf(b)} $\Delta = 1.5$ and $\Delta_2 = 1.5$. We have $L = 24$ in all cases, with the average of 6 random-state realizations shown per choice of parameters. Dashed lines are Gaussian fits. To reduce finite-site effects, the 5 sites furthest from the central site are not shown here and are not used in the Gaussian fits. Error bars are $\pm2$ standard errors of the mean (assuming Gaussian errors).}
\label{Fig::Transport_2}
\end{figure}

To proceed, we can study the emerging hydrodynamics in more detail by analyzing the full spatial profile $C(r,t)$ at a fixed time $t = 9$ in Fig.\ \ref{Fig::Transport_2}, where we again consider the integrable $\Delta_2 = 0$ chain [Fig.\ \ref{Fig::Transport_2}~(a)] and a nonintegrable $\Delta = \Delta_2 = 1.5$ parameter choice [Fig.\ \ref{Fig::Transport_2}~(b)]. In the Hermitian $g = 0$ model, $C(r,t)$ is known to be well described by Gaussians signalling normal diffusion \cite{Bertini_2021, Richter_2018}. 
Consistent with the faster decay of $C(0,t)$ in Fig.\ \ref{Fig::Transport_1}, we find that the profiles $C(r,t)$ are broader for $g > 0$ compared to the Hermitian reference case. Interestingly, at least in the bulk of the system where finite-size effects are less relevant, $C(r,t)$ appears to remain approximately Gaussian in the non-Hermitian model. In this context, let us also emphasize that, even though a finite $g > 0$ leads to an asymmetry between left and right hopping amplitudes, the spreading of correlations measured by  $C(r,t)$ remains symmetric. This is due to the fact that we here consider the ``grand-canonical ensemble'', i.e., we average $C(r,t)$ over all subsectors of total magnetization $M^z$.  

\subsubsection{Spin chain with non-Hermitian disorder}

\begin{figure}[tb]
\centering
\includegraphics[width=\columnwidth]{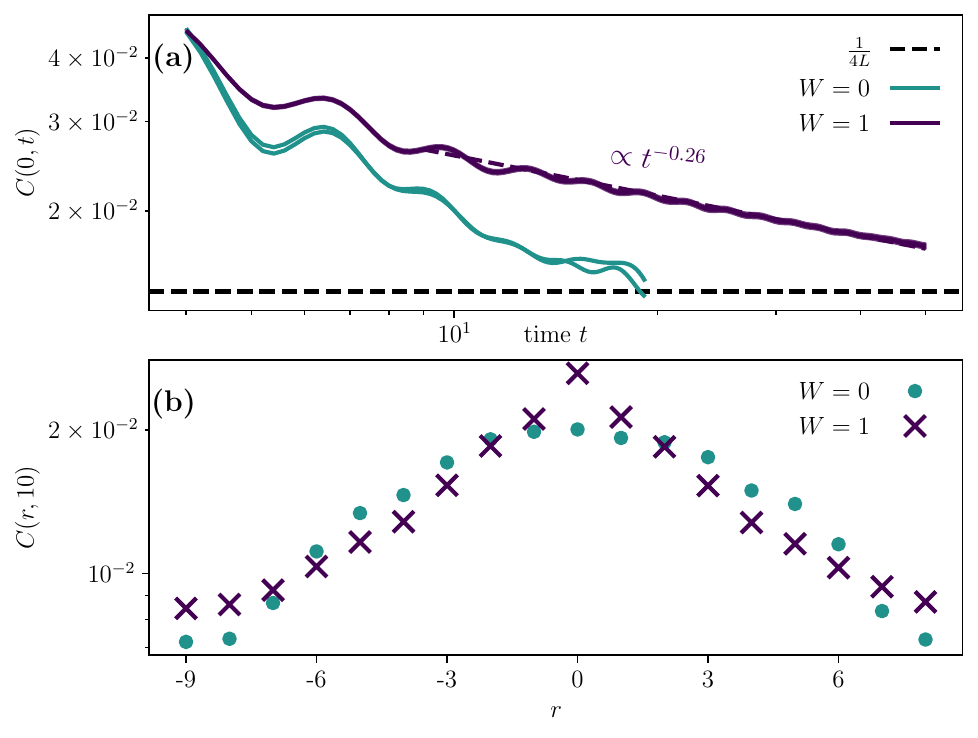}
\caption{\textbf{(a)} Spin-Spin autocorrelation function $C(0,t)$ for $L = 18$ and $W = 0,1$, obtained using DQT. The $W=1$ curve is the average of $\sim 10^4$ random-state und disorder realizations, and the area within $\pm2$ standard errors of the mean (assuming Gaussian errors) is shaded. A power-law fit (starting at $t = 9$) to the $W = 1$ hydrodynamic tail in shown as well. The $W = 0$ curve shows data from two random-state realizations, which can be seen to agree almost perfectly. The horizontal dashed line indicates the expected long-time value of $C(0,t)$. \textbf{(b)} Spatial profiles for same data as above showing snapshot at $t = 10$. As can be seen from the semilogarithmic plot, $C(r,t)$ decays approximately exponentially with $r$ in the disordered model.}
\label{Fig::Random_Potential_Transport}
\end{figure}

We now study the model given in Eq.\ \eqref{Eq::HDisorder}, i.e., an XXZ chain perturbed by a non-Hermitian random potential. In Fig.\ \ref{Fig::Random_Potential_Transport}~(a), the on-site correlation function $C(0,t)$ is shown for $W = 0$ and $W = 1$ with $L = 16,18$. As a point of reference, superdiffusive $\propto t^{-2/3}$ scaling of $C(0,t)$ is expected in the Hermitian $W = 0$ case \cite{Bulchandani_2021, Gopalakrishnan_2023, Bertini_2021, Ljubotina_2019}. In contrast, for the  non-Hermitian chain with $W = 1$, we find that $C(0,t)$ decays notably slower, but is consistent with eventual thermalization in the weakly disordered regime \cite{Roccati2023} as it appears to approach the equilibrium value $1/(4L)$ at long times.

The full spatial profile $C(r,t)$ at fixed time $t = 10$ is shown in Fig.\ \ref{Fig::Random_Potential_Transport}~(b). We find that the presence of disorder has a clear effect on the spreading of correlations. In particular, $C(r,t)$ decays approximately exponentially with $r$ for $W = 1$. This slow anomalous dynamics of $C(r,t)$ is akin to the spreading of correlations in the putative subdiffusive regime of the standard Hermitian MBL model below the localization transition \cite{Bera_2017, Richter_2018_2, Luitz_2017}. While it is challenging to numerically confirm the asymptotic scaling form of $C(r,t)$ at very long times and distances, Fig.\ \ref{Fig::Random_Potential_Transport}~(b) shows that the non-Hermitian disorder has a qualitatively stronger impact on the shape of $C(r,t)$ compared to the non-reciprocal hopping studied in Fig.\ \ref{Fig::Transport_2}. In the future, it would also be interesting to study the transport properties and the dynamics of Eq.\ \eqref{Eq::HDisorder} at stronger disorder strengths, where the level-spacing statistics indicate a crossover from chaotic to localized behavior \cite{Roccati2023}.   
%


\section{Conclusion}\label{Sec::Conclu}

The diverse properties of non-Hermitian Hamiltonians have recently attracted increased attention in the context of nonequilibrium many-body quantum dynamics. In this paper, we have studied the impact of non-Hermiticity on transport properties and integrability in systems with a global conservation law. More specifically, we have considered different non-Hermitian perturbations to the one-dimensional spin-$1/2$ XXZ chain and analyzed the hydrodynamic scaling of time-dependent spin-spin correlation functions. We have proposed a generalization of such dynamical correlation functions to the case of non-unitary time evolution with a straightforward experimental interpretation. 

For the asymmetric XXZ chain (i.e., Hatano-Nelson model), which is non-Hermitian due to the non-reciprocal hopping amplitudes, we observed a crossover from diffusive to fast (seemingly ballistic) transport even for rather weak values of the non-Hermitian perturbation. Interestingly, the signatures of ballistic transport emerged even in the presence of next-nearest neighbor interactions for which the model becomes nonintegrable. This finding is surprising as it contrasts the usual expectation that chaotic quantum systems with short-ranged interactions show diffusive transport. We also considered a spin chain subjected to a non-Hermitian disorder potential, which may undergo a transition to a non-Hermitian many-body localized phase at strong disorder \cite{Roccati2023}. Here, we focused on moderate disorder strengths for which the model is chaotic (Appendix \ref{Sec::App_LSDisorder}) and observed seemingly thermalizing, yet anomalous transport with a non-Gaussian spreading of spatiotemporal correlations.

From a technical point of view, we obtained our numerical results by exploiting the concept of dynamical quantum typicality. Even though the non-unitary time evolution reduces the accuracy of DQT at long times, we have shown that DQT still allows to accurately simulate correlation functions on intermediate time scales for system sizes beyond the range of standard exact diagonalization. While random states and DQT are by now well established numerical tools, this work represents, to the best of our knowledge, the first application of DQT to the dynamics of non-Hermitian many-body systems (but see \cite{Heitmann_2023, Heitmann_2023_2} for related ideas).   

A natural direction of future research is to study non-unitary quantum dynamics and transport in a wider range of models, such as non-Hermitian models with higher-order multipole conservation laws \cite{Gliozzi2024}. It would also be interesting to better understand the properties of off-diagonal matrix elements entering the eigenstate thermalization hypothesis in non-Hermitian many-body quantum systems \cite{Roy2023}, including their effect on dynamics, transport, and thermalization.

The code used to generate the data in this paper can be found at the following repository: \url{https://github.com/DylanMahoney/NonHermitianSystems}.

\subsection*{Acknowledgements}
We thank Vedika Khemani, Yaodong Li, J\'ozsef M\'ak, and Robin Steinigeweg for helpful discussions. 
D.\,E.\,M.\, acknowledges support from a Vice Provost for Undergraduate Education Major Grant from Stanford University.
J.\,R.\, is funded by the European Union's Horizon Europe research 
and innovation programme, Marie Sk\l odowska-Curie grant no.\ 101060162, and by the Packard Foundation
through a Packard Fellowship in Science and Engineering (V.\ Khemani's grant). 


\appendix

\section{Integrability-breaking in the XXZ chain perturbed by random imaginary fields}\label{Sec::App_LSDisorder}

\begin{figure}[tb]
\centering
\includegraphics[width=\columnwidth]{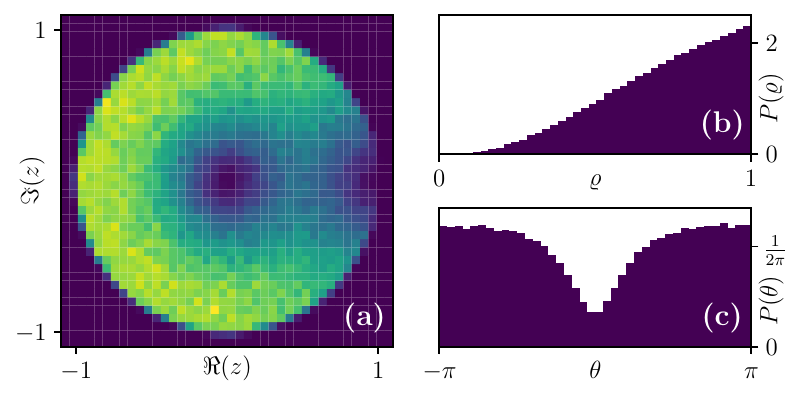}
\caption{Level-spacing statistics for the model in Eq.\ \eqref{Eq::HDisorder} with $W = 1$ and $L = 14$ in the $M^z = 1$ magnetization sector, obtained from 100 random disorder realizations. {\bf(a)} Distribution of  $z = \varrho e^{i \theta}$ on the complex plane  [Eq.~\eqref{Eq::zm}]. {\bf [(b), (c)]} Marginal distributions $P(\varrho)$ and $P(\theta)$.}
\label{Fig::LevelStatRandomPotential}
\end{figure}

Analogous to Fig.\ \ref{Fig::LevelStatHN_1} in the main text, Fig.\ \ref{Fig::LevelStatRandomPotential} shows results for the level statistics of the disordered non-Hermitian Hamiltonian in Eq.\ \eqref{Eq::HDisorder}. The data are obtained in the $M^z = 1$ subspace and are accumulated over 100 random disorder realizations using a moderate value of $W = 1$. As can be clearly seen in Fig.\ \ref{Fig::LevelStatRandomPotential}, the distribution $P(z)$ is inhomogeneous with suppressed weight around $\varrho= 0$ and $\theta = 0$. Thus, in contrast to the non-Hermitian hopping asymmetry $g \neq 0$ in Eq.\ \eqref{Eq::HN}, which did not break the integrability of the XXZ chain, the random non-Hermitian potential appears to render the model in Eq.\ \eqref{Eq::HDisorder} chaotic and nonintegrable.

\section{General discussion of correlation functions for non-Hermitian systems}\label{Sec::App_Correl}

Let us provide further details on the generalization of correlation functions to non-Hermitian systems. In the following, we will restrict ourselves to correlation functions at formally infinite temperature, i.e., evaluated with respect to the maximally-mixed state $\rho_{\infty} = \mathbbm{1}/(\text{dim} \mathcal{H})$, where $\mathcal{H}$ denotes the Hilbert space of the system. 

Thus if $A$ and $B$ are Hermitian operators $\mathcal{H} \to \mathcal{H}$ such that $[A(t),B] = 0$ for all $t \geq 0$ in the Heisenberg picture with a Hermitian Hamiltonian, we can define the infinite-temperature correlation function as,
\begin{equation}\label{Eq::HermitianITC}
   C(t) \equiv \braket{A(t)B}_{\infty} = \text{tr}[A(t)B\rho_{\infty}]\ , 
\end{equation}
i.e., the expectation value of the Heisenberg operator $A(t)B$ given density matrix $\rho_{\infty}$. Note that we here assume, without loss of generality, that $\braket{A}_{\infty} = \braket{B}_{\infty}= 0$.

Generalizing this to the non-Hermitian case is challenging due to the definition of non-unitary time evolution in Eq.~\eqref{Eq::time_evolution_for_rho}. In particular, it is unclear how to meaningfully define the Heisenberg-evolved operator $A(t)$ in the non-Hermitian case.

However, we can find another expression for the infinite-temperature correlator of $A$ and $B$, which agrees with Eq.\ \eqref{Eq::HermitianITC} for Hermitian Hamiltonians, but remains applicable for non-Hermitian Hamiltonians as well.
To this end, we decompose the operator $B$ into $B = \sum_{b} b\Pi_{B = b}$, where each $\Pi_{B = b}$ is a projector onto the $B = b$ eigenspace. Then, using the linearity of the trace, we can write, 
\begin{equation}
    \text{tr}[A(t)B\rho_{\infty}] = \sum_b b\text{tr}[A(t)\Pi_{B = b}\rho_{\infty}]\ .
\end{equation}
Note that for Hermitian Hamiltonians, using the cyclic property of the trace, this is equivalent to,
\begin{equation}
    C(t) = \sum_b b\text{tr}[Ae^{-iHt}\Pi_{B = b}\rho_{\infty}e^{iHt}]\ . 
\end{equation}
Let $\rho_{b = B}$ now be the density matrix of the system starting at infinite temperature, immediately after measuring $B$ to be $b$. Then $\rho_{B = b} = \Pi_{B = b}\rho_{\infty} \Pi_{B = b}/\text{tr}[\Pi_{B = b}\rho_{\infty}]$, so $C(t)$ can be rewritten (where we also used the projection property $\Pi_{B = b}^2 = \Pi_{B = b}$),
\begin{equation}
C(t) = \sum_b b \text{tr}[\Pi_{B = b}\rho_{\infty}]\text{tr}[Ae^{-iHt}\rho_{B = b}e^{iHt}]\ , 
\end{equation}
or, in the Schr\"odinger picture,
\begin{equation}\label{Eq::generalized_ITC_no_M}
  C(t) =  \sum_b b \text{tr}[\Pi_{B = b}\rho_{\infty}]\text{tr}[A\rho_{B = b}(t)]\ .
\end{equation}
This expression can be generalized to non-Hermitian Hamiltonians via Eq.\ \eqref{Eq::time_evolution_for_rho}. If $B$ commutes with a third observable, say the total magnetization $M^z$ with eigenvalues $m$, similar algebra allows us to (equivalently in the Hermitian case) write the correlator of $A$ and $B$ as,
\begin{equation}\label{Eq::generalized_ITC_with_M}
    C(t) = \sum_{b,m} b \text{tr}[\Pi_{B = b,M^z = m}\rho_{\infty}]\text{tr}[A\rho_{B = b,M^z = m}(t)]\ ,
\end{equation}
where we have used that $M^z$ and $B$ have simultaneous eigenspaces, and where $\rho_{B = b,M^z = m}$ is the density matrix of the system if it starts in density matrix $\rho_{\infty}$ and measurements are made of $B$ and $M^z$ yielding results $b$ and $m$.

Equation \eqref{Eq::generalized_ITC_no_M} can be interpreted as the outcome of the following procedure: Measure $B$ at time $0$ and measure $A$ at time $t$, and over many runs compute the expected value of $A$ at time $t$ conditional on the different possible outcomes for $B$, and the probability of each measurement outcome for $B$. Equation \eqref{Eq::generalized_ITC_with_M} can be interpreted as following the same procedure, except also measuring $M^z$ at time 0 and keeping track of the expected value of $A$ given the different possible outcomes for $B$ and $M^z$ simultaneously.

Equations \eqref{Eq::generalized_ITC_no_M} and \eqref{Eq::generalized_ITC_with_M} agree in the Hermitian case, but they do not always agree in the case of non-Hermitian Hamiltonians due to the non-unitary time evolution (see below). In this paper, using $A = S^z_{\ell}$, $B = S^z_{L/2}$, we will take Eq.\ \eqref{Eq::generalized_ITC_with_M} as our definition of the infinite-temperature correlator so that total spin $M^z$ is conserved, thereby enabling meaningful discussion of transport properties.

\subsubsection*{Details on the disagreement of Eqs.\ \eqref{Eq::generalized_ITC_no_M} and \eqref{Eq::generalized_ITC_with_M} in non-Hermitian systems}

A key aspect is that the time evolution of density matrices in non-Hermitian systems [Eq.\ \eqref{Eq::time_evolution_for_rho}] becomes nonlinear. Consider, for instance, a two-level system with eigenvectors $\ket{0}$ and $\ket{1}$ with eigenvalues $+i$ and $-i$, respectively. The density matrices $\rho_0 = \ket{0}\bra{0}$ and $\rho_1 = \ket{1}\bra{1}$ are both constant in time using Eq.\ \eqref{Eq::time_evolution_for_rho}, which means that if time evolution were linear, the density matrix $\rho_{\text{mix}} = (\rho_0 + \rho_1)/2$ would also be constant in time. But we can compute that,
\begin{align}
    \frac{e^{-iHt}\rho_{\text{mix}}e^{i H^{\dag}t}}{\text{tr}[e^{-iHt}\rho_{\text{mix}}e^{i H^{\dag}t}]} & = \frac{\frac 1 2 e^{2t}\ket{0}\bra{0} + \frac 1 2 e^{-2t}\ket{1}\bra{1}}{\text{tr}[\frac 1 2 e^{2t}\ket{0}\bra{0} + \frac 1 2 e^{-2t}\ket{1}\bra{1}]} \nonumber\\
    & = \frac{e^{2t}}{2\cosh(2t)}\rho_0 + \frac{e^{-2t}}{2\cosh(2t)}\rho_1
\end{align}
is not constant in time.

A more concrete example of the disagreement between Eqs.\ \eqref{Eq::generalized_ITC_no_M} and \eqref{Eq::generalized_ITC_with_M} for one of our models can be seen analytically for the Hamiltonian defined in Eq.\ \eqref{Eq::HDisorder} when $L = 2$, $h_1 = 2$, and $h_2 = 0$, where $B = S^z_1$ and $A = M^z = S^z_1 + S^z_2$. (Note that Eq.\ \eqref{Eq::generalized_ITC_no_M} and Eq.\ \eqref{Eq::generalized_ITC_with_M} are both linear in $A$, so setting $A = M^z$ is the same as first setting $A = S^z_1$, then setting $A = S^z_2$, and adding the two results.) In this case the matrix expressions appearing in Eqs.\ \eqref{Eq::generalized_ITC_no_M} and \eqref{Eq::generalized_ITC_with_M} are simple enough to be evaluated symbolically, and we get,
\begin{equation}
    C_{\eqref{Eq::generalized_ITC_no_M}}(t) = \frac{1}{4}-\frac{t^4}{4 t^4+\left(4 t^2+2\right) \cosh (2 t)-4 t \sinh (2 t)+2}\ ,
\end{equation}
while,
\begin{equation}
    C_{\eqref{Eq::generalized_ITC_with_M}}(t) = \frac{1}{4}\ .
\end{equation}
Thus we see that Eq.\ \eqref{Eq::generalized_ITC_with_M} conserves magnetization, whereas Eq.\ \eqref{Eq::generalized_ITC_no_M} does not.

\section{Derivation of DQT relation}\label{Sec::App:Typ}

In this section, we provide a derivation of the pure-state approximation $C_\psi(r,t)$ [Eq.\ \eqref{Eq::TypCrt} in main text]. To this end, we start from the expression of $C(r,t)$ in Eq.\ \eqref{Eq::Correl_NonHermitian},
\begin{equation}
    C(r,t) = \sum_{s,m} s\ \text{tr}[\Pi_{s,m}\rho_\infty]\ \text{tr}[S_{\ell+r}^z \rho_{s,m}(t)]\ . \label{Eq::Correl_NonHermitian_In_Appendix}
\end{equation}
Recall that $\rho_{s,m} = \Pi_{s,m}\rho_\infty\Pi_{s,m}/\text{tr}[\Pi_{s,m}\rho_\infty]$ and $\rho_\infty = \mathbb{1}/\text{tr}[\mathbb{1}]$.
Now we use Eq.\ \eqref{Eq::time_evolution_for_rho} and the cyclic invariance of the trace to rewrite $\text{tr}[S_{\ell+r}^z \rho_{s,m}(t)]$ as, 
\begin{align}
    \text{tr}[S_{\ell+r}^z \rho_{s,m}(t)] &= \frac{\text{tr}[S_{\ell+r}^z e^{-iHt}\Pi_{s,m}\rho_\infty e^{iH^\dagger t}]}{\text{tr}[e^{-iHt}\Pi_{s,m}\rho_\infty e^{iH^\dagger t}]} \nonumber \\
    &=\frac{\text{tr}[\Pi_{s,m} e^{iH^\dagger t}S_{\ell+r}^z e^{-iHt}\Pi_{s,m}]}{\text{tr}[\Pi_{s,m} e^{iH^\dagger t} e^{-iHt} \Pi_{s,m}]} \nonumber \\
    &\approx  \frac{\bra{\psi}\Pi_{s,m} e^{iH^\dagger t}S_{\ell+r}^z e^{-iHt}\Pi_{s,m}\ket{\psi}}{\bra{\psi}\Pi_{s,m} e^{iH^\dagger t} e^{-iHt} \Pi_{s,m}\ket{\psi}}  \nonumber \\
    &= \bra{\psi_{s,m}(t)}S_{\ell+r}^z\ket{\psi_{s,m}(t)}\ , \label{Eq::Dev1}
\end{align}
where we have used DQT to approximate the trace by a normalized Haar-random pure state $\ket{\psi}$, and exploited the projection property $\Pi_{s,m}^2 = \Pi_{s,m}$. Moreover, in the last step, we have used the definition of $\ket{\psi_{s,m}(t)}$ [Eq.\ \eqref{Eq::RandomState}] together with the definition of non-unitary time evolution [Eq.~\eqref{Eq::Psit}]. Plugging Eq.\ \eqref{Eq::Dev1} into Eq.\ \eqref{Eq::Correl_NonHermitian_In_Appendix}, we recover Eq.~\eqref{Eq::TypCrt} as desired.

\section{The inverse participation ratio in non-Hermitian systems}\label{Sec::App_IPR}

Let us provide further motivation on the choice of Eq.\ \eqref{Eq::IPR} as our definition of the inverse participation ratio in non-Hermitian systems. We start with the usual definition of the inverse participation ratio in Hermitian systems, already given in Eq.\ \eqref{Eq::Hermitian_IPR},
\begin{equation}\label{Eq::Hermitian_IPR_in_appendix}
    I(t) = \sum_{n=1}^{{\cal{N}}} |\braket{\phi_n|\psi(t)}|^4\ ,
\end{equation}
where ${\cal{N}}$ is the Hilbert space dimension.
For any state $\ket{\phi_n}$, let $A_n$ now be the Hermitian operator on the Hilbert space defined by
\begin{equation}
    \begin{cases}
        A_n\ket{\phi_n} = \ket{\phi_n}\\
        A_n\ket{v} = -\ket{v} & \ket{v} \in (\text{span}(\ket{\phi_n}))^{\perp}
    \end{cases}\ ,
\end{equation}
where $(\text{span}(\ket{\phi_n}))^{\perp}$ is the orthogonal complement of the span of $\ket{\phi_n}$. Then we can rewrite Eq.\ \eqref{Eq::Hermitian_IPR_in_appendix} as
\begin{equation}
    I(t) = \sum_{n = 1}^{{\cal{N}}} \text{Pr}[\text{a measurement of $A_n$ yields $+1$}]^2\ .
\end{equation}
Now let us define a measurement procedure which we will refer to as $\mathcal{P}$:
\begin{enumerate}
    \item Choose a uniform random number $k$ from $\{1,\dots,\cal{N}\}$.
    \item Measure $A_k$.
\end{enumerate}
If the outcome of Step 2 is +1, then we define the ``output'' of $\mathcal{P}$ to be $k$. Otherwise, we define the ``output'' of $\mathcal{P}$ to be ``failure''. Then we have
\begin{equation}\label{eq::IPR_from_probabilities}
    I(t) = \sum_{n = 1}^{{\cal{N}}} \text{Pr}[\text{$\mathcal{P}$ outputs $n$}\ | \ \text{$\mathcal{P}$ doesn't fail}]^2\ ,
\end{equation}
because
\begin{align}
    & \text{Pr}[\text{$\mathcal{P}$ outputs $n$}\ | \ \text{$\mathcal{P}$ doesn't fail}]\label{Eq::probability_manipulation}\\
    & = \frac{\text{Pr}[\text{$\mathcal{P}$ outputs $n$}]}{\text{Pr}[\text{$\mathcal{P}$ doesn't fail}]}\nonumber\\ 
    & = \frac{\text{Pr}[\text{$\mathcal{P}$ outputs $n$}]}{\sum_{m = 1}^{{\cal{N}}}\text{Pr}[\text{$\mathcal{P}$ outputs $m$}]}\nonumber\\
    & = \frac{\text{Pr}[k = n]\text{Pr}[\text{a measurement of $A_n$ yields $+1$}]}{\sum_{m = 1}^{{\cal{N}}} \text{Pr}[k = m]\text{Pr}[\text{a measurement of $A_m$ yields $+1$}]}\nonumber\\
    & = \frac{1\cal{N}}{1/\cal{N}}\frac{\text{Pr}[\text{a measurement of $A_n$ yields $+1$}]}{\sum_{m = 1}^{{\cal{N}}} |\braket{\phi_m|\psi(t)}|^2}\nonumber\\
    & = \frac{\text{Pr}[\text{a measurement of $A_n$ yields $+1$}]}{\braket{\psi|\psi}}\nonumber\\
    & = \text{Pr}[\text{a measurement of $A_n$ yields $+1$}]\nonumber\ ,
\end{align}
where we have used that $\{\phi_m\}$ form a complete basis for the Hilbert space. Let us now turn to the case of non-Hermitian systems, where we set $\{\ket{\phi_k}\}_{k = 1}^{{\cal{N}}}$ to be the normalized right eigenvectors of the Hamiltonian, either in the whole Hilbert space or in a fixed symmetry subspace. The measurement procedure $\mathcal{P}$ remains perfectly well-defined for this choice of $\{\ket{\phi_k}\}_{k = 1}^{{\cal{N}}}$, so let us define $I(t)$ for non-Hermitian systems to be given by Eq. \eqref{eq::IPR_from_probabilities}. The only notable complication comes from the fact that the right eigenvectors cannot be assumed to be an orthonormal basis for the Hilbert space, in which case the reasoning of Eq.\ \eqref{Eq::probability_manipulation} can be slightly modified to yield
\begin{equation}\label{Eq::normalization_source}
    \text{Pr}[\text{$\mathcal{P}$ outputs $n$}\ | \ \text{$\mathcal{P}$ doesn't fail}] = \frac{|\braket{\phi_n|\psi(t)}|^2}{\sum_{m = 1}^{{\cal{N}}} |\braket{\phi_m|\psi(t)}|^2}\ .
\end{equation}
Plugging Eq.\ \eqref{Eq::normalization_source} into the definition of $I(t)$ in Eq.\ \eqref{eq::IPR_from_probabilities} yields the definition of $I(t)$ given in Eq.\ \eqref{Eq::IPR}.

\section{Current-current correlation functions}\label{Sec::App::CurrCurr}

In addition to spin-spin correlation functions, we can also consider current correlation functions ${\cal J}(t) = \text{tr}[ J(t) J \rho_\infty]$, 
which we write in a form analogous to Eq.\ \eqref{Eq::Correl_NonHermitian} that is suitable for non-Hermitian systems with non-unitary time evolution, 
\begin{equation}\label{Eq::Curr::Corr}
    {\cal J}(t) = \sum_{j,m} j\ \text{tr}[\Pi_{j,m}\rho_{\infty}]\langle J(t)\rangle_{\Pi_{j,m}}\ .
\end{equation}
Here, we have again exploited that $M^z$ and the current $J = \sum_j j \Pi_j$ commute with each other. Specifically, we consider the well-known Hermitian spin-current of the XY or XXZ chain \cite{Bertini_2021}, 
\begin{equation}\label{Eq::CurrentOp}
    J = \frac{i}{2}\sum_{\ell=1}^L \left(S_\ell^+ S_{\ell+1}^- - S_\ell^- S_{\ell+1}^+\right)\ .
\end{equation}
\begin{figure}[b]
\centering
\includegraphics[width=\columnwidth]{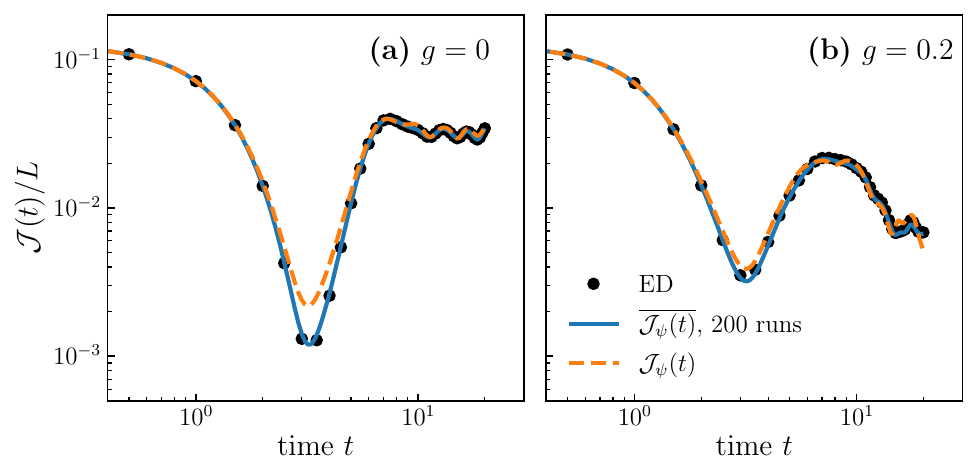}
\caption{Comparison between dynamical quantum typicality and exact diagonalization for system size $L = 10$, analogous to Fig.\ \ref{Fig::Typicality_1} in the main text, but now for the current correlation function ${\cal J}(t)$. Data are shown for $\Delta = 1.5$, $\Delta_2 = 0$ with {\bf (a)} $g = 0$ and {\bf (b)} $g = 0.2$.}
\label{Fig::Current_Compare}
\end{figure}
The current operator $J = \sum_\ell J_\ell$ is obtained from a lattice continuity equation $\tfrac{d}{dt} S_{\ell}^z = i[H_{XXZ},S_\ell^z] = J_{\ell-1}-J_\ell$ \cite{Bertini_2021}. Since $[S_{\ell'}^z,{S_\ell}^z] = 0$, the expression of $J$ in \eqref{Eq::CurrentOp} remains unchanged for the disordered non-Hermitian chain in Eq.\ \eqref{Eq::HDisorder}. On the other hand, for the Hatano-Nelson model \eqref{Eq::HN}, the continuity equation would actually yield a non-Hermitian current operator with hopping asymmetry. We here decide to still study the form given in Eq.\ \eqref{Eq::CurrentOp} in order to keep operators Hermitian (except for $H$ of course). Another reason for studying ${\cal J}(t)$ with the Hermitian version of $J$ \eqref{Eq::CurrentOp} is given by the fact that, as mentioned in Sec.\ \ref{Sec::Model}, the asymmetric hopping terms in Eq.\ \eqref{Eq::HN} can be obtained as $\cosh(g) H_{XY} + i \sinh(g) J$. The non-Hermiticity of Eq.\ \eqref{Eq::HN} can thus be interpreted as an external driving by the current $J$ \cite{Panda_2020}. 

DQT expressions similar to Eqs.\ \eqref{Eq::TypCrt} and \eqref{Eq::RandomState} can be obtained also for a pure-state approximation ${\cal J}_\psi(t)$ of the current correlation function ${\cal J}(t)$. In Fig.\ \ref{Fig::Current_Compare}, a comparison of DQT and ED is shown using a small system size $L = 10$. Analogous to Fig.\ \ref{Fig::Typicality_1} in the main text, we observe a convincing agreement between the averaged pure-state approach and the exact result. 
Let us note, however, that in contrast to the spin-spin correlation function, it is less obvious that DQT yields a computational advantage when considering ${\cal J}(t)$. 
Namely, while the local spin operator $S_\ell^z$ is diagonal in the computational basis, such that the projection $\Pi_{s,m}$ in the case of $C(r,t)$ can be easily applied, the projection $\Pi_{j,m}$ in Eq.\ \eqref{Eq::Curr::Corr} requires diagonalization of $J$. Thus, the system sizes reachable using the pure-state approach will be smaller than in the case of $C(r,t)$. However, since a single diagonalization of $J$ is not necessarily the most time-consuming part of the simulation, DQT might still allow to study slightly larger systems than accessible by standard ED.

\end{document}